\documentclass{article} 
\usepackage{\jobname,epsfig,amsmath,amssymb} 
 
\newcommand{\sezione}[2]{
\refstepcounter{section}\label{#2}
\setcounter{equation}{0} 
\setcounter{subsection}{0} 
\addcontentsline{toc}{section}
      {\normalsize\textbf{\thesection.\ #1}} 
\bigskip\bigskip\noindent 
\normalsize\textbf{\thesection.\ #1}\smallskip\nopagebreak}
\def\thesection{{\normalsize\arabic{section}}} 
\newcommand{\subsec}[2]{
\refstepcounter{subsection}\label{#2}
\addcontentsline{toc}{subsection}
      {\normalsize\normalfont\textit{\thesubsection.\ #1}} 
\medskip\medskip\noindent 
\normalsize\normalfont\textit{\thesubsection.\ #1}\smallskip\nopagebreak}
\def\thesubsection{{\normalsize\normalfont
            \textit{\arabic{section}.\arabic{subsection}}}} 

\def\rect#1#2{{\vcenter{\vbox{\hrule height.3pt 
            \hbox{\vrule width.3pt height#2truecm \kern#1truecm 
            \vrule width.3pt} 
            \hrule height.3pt}}}}

\def\aversim#1#2{\lower3pt\vbox{\baselineskip0pt \lineskip-.1pt 
    \ialign{$\mathsurround=0pt #1\hfil##\hfil$\crcr#2\crcr\sim\crcr}}}

\newcommand{\rh}[1]{\rho_{\scriptscriptstyle{#1}}}
\newcommand{\si}[1]{\sigma_{\scriptscriptstyle{#1}}}
\newcommand{\la}[1]{\lambda_{\scriptscriptstyle{#1}}}
\newcommand{\Tr}[1]{{\mathrm{Tr}}_{\scriptscriptstyle{#1}}}

\begin{document} 
\title{Folding transitions of the square--diagonal two--dimensional
lattice} 
\author{ 
\\ \\ 
\normalsize{Emilio N.M. Cirillo}\\ 
\normalsize{Dipartimento di Matematica, 
II Universit\`a di Roma -- Tor Vergata,} \\ 
\normalsize{via della Ricerca Scientifica, I--00133 Roma, Italy.}
\\ \\
\normalsize{Giuseppe Gonnella}\\ 
\normalsize{Istituto Nazionale per la Fisica della Materia, Unit\`a di Bari}\\
\normalsize{and Dipartimento di Fisica dell'Universit\`{a} di Bari}\\
\normalsize{and Istituto Nazionale di Fisica Nucleare, Sezione di Bari}\\
\normalsize{via Amendola 173, 70126 Bari, Italy.}
\\ \\ 
\normalsize{Alessandro Pelizzola}\\ 
\normalsize{Dipartimento di Fisica del Politecnico di Torino}\\
\normalsize{and Istituto Nazionale per la Fisica della Materia,}\\ 
\normalsize{c. Duca degli Abruzzi 24, 10129 Torino, Italy.}
} 
\date{} 
\maketitle 
\vskip 2 cm 
\begin{abstract} 
The phase diagram of a vertex model introduced by P. Di Francesco 
(Nucl. Phys. B {\bf 525}, 507 1998) representing the configurations of a 
square
lattice which can fold with different bending energies
along the main axes and the diagonals has been studied 
by Cluster Variation Method. 
A very rich structure with partially and completely folded phases,
different disordered phases and a flat phase is found.
The crumpling transition between a disordered and  the flat phase 
is first--order. The CVM 
results are confimed by the analysis of the ground states and of the
two limits where the model reduces to an Ising model.
\end{abstract} 
 
\newpage 
\sezione{Introduction}{sec:intr}
\par\noindent 
Polymerized membranes are two--dimensional networks of molecules with
fixed connectivity \cite{NPW}.  Their configurational properties and
the existence of a crumpling transition between a folded and a flat
phase are relevant for the behavior of some biological systems
\cite{NP}.  Models of polymerized membranes can be defined on regular
lattices.  The constraint of fixed connectivity gives rise to the
definition of complicate vertex models which offer the advantage that
explicit analytical calculations can be performed \cite{Bo}.  In these
models the bonds between the vertices of the network have a fixed
length and also, as a further simplification, self--avoidance is not
considered.  In this way the only degrees of freedom of the network
are related to its possible states of folding and each state can be
weighted by a Boltzmann factor depending on the relative angle between
adjacent plaquettes in the network.
 
The model studied in this paper has been introduced by Di Francesco
\cite{Di1,Di2} and concerns the folding properties of a square lattice
which can be folded along the main axes and along the diagonals.  This
model follows previous studies of the folding properties of a
triangular network embedded in $d$--dimensional lattices. In the case
$d=2$ \cite{KJ, epl} each triangle of the network can be only in a 
``up'' or ``down'' state. The entropy of this problem is the
same as in the three--colouring problem of the honeycomb lattice
\cite{B}.  The introduction of a bending rigidity induces a
first--order crumpling transition which has been studied in
\cite{pre,nostro}. The model with the triangular network embedded in
the face--centered--cubic lattice was defined in \cite{npb95} and a
first--order transition has been found also in this case
\cite{nostro1,Gu}.
 
In this paper we consider the two--dimensional folding of the
square--diagonal lattice and study the phase diagram of this system in
terms of the bending rigidities $\kappa_S$ and 
$\kappa_L$ relative to short and long edges.  Actually, the phase
behavior of this model was already considered in \cite{Di2} by means
of a transfer matrix analysis, but the ground state was not properly
investigated, and as a consequence some phases were missing. Here we
consider the complete set of the ground states of the model and study
the phase diagram at finite temperatures by the cluster variation
method (CVM) \cite{Kik1,An,Morita,ptps}.  This method has already
proven to be useful for studying vertex models and folding problems
\cite{nostro,nostro1,Gu}.
 
The outline of the paper is the following. In the next Section we
define the model; the ground states will be shown in Section 3.  In
Section 4 the CVM approximation used in this paper will be described
and our results will be presented in Section 5.
 
\sezione{The square--diagonal folding model}{sec:modello}
\par\noindent 
We consider the folding configurations of the square--diagonal lattice
where any couple of adjacent triangles can be on the same plane or
with the two triangles one on the other. Triangles can be folded with
respect to the main axes of the square lattice or with respect to the
diagonals so that in the definition of the bending energy different
couples of triangles with a short edge or a long edge in common have
to be considered. We will start the definition of the model by
enumerating the folding states in terms of loop configurations.

\subsec{Folding and loop configurations}{sub:modello} 
\par\noindent 
First fix a reference orientation for all the edges of the lattice in
such a way that for each triangle the vectorial sum of the oriented
edges is zero. There are two choices for this global orientation and
one is shown in Fig. \ref{fig:uno}.  In any folding configuration each
short edge is mapped on one of the four vectors $\pm \vec e_1, \pm
\vec e_2$; this mapping defines the state of the network.  Then
consider two triangles with a long edge in common as the pair shaded
in Fig. \ref{fig:uno}. This pair can be in two folding states that can
be represented in the following way: consider another square lattice
as in Fig. \ref{fig:due}.  The lines joining the centers of the edges
are dashed or full depending on whether they are dual to the mapped
short edge vectors $\pm \vec e_1$ or $\pm \vec e_2$ respectively.  The
obvious observation that in each triangle there is a short edge vector
in the direction of $\pm \vec e_1$ and an edge in the direction of
$\pm\vec e_2$ implies that for each center of the edges of the new
square lattice there will be two dashed and two full lines. As a
consequence the dashed lines and the full lines will form two sets of
orthogonal closed loops \cite{Di1}.
 
Taking into account also the orientation of $\vec e_1$ and $\vec e_2$,
it comes out that each closed loop has its own orientation
independently from the others. Therefore the original folding problem
is equivalent to a dense two--loop problem where a sign representing
the orientation has to be attributed to each closed loop. Further
details can be found in \cite{Di1}.
 
Consider now the square plaquettes of this new lattice. In each
plaquette there are two dashed and two full lines. On each line there
can be two possible signs so that there are $2\times 2^4 = 32 $ states
for each plaquette.  Of course, the sign conservation law along the
closed loops implies that only some of the $32^2 $ configurations of
two neighboring plaquettes are allowed.  The logarithm of the number
of all possible states of these plaquette configurations gives the
entropy of the folding problem.

\subsec{Bending energy}{sub:bending} 
\par\noindent 
Here we introduce a bending energy 
weighting the different folding configurations. 
We attribute a Boltzmann weight 
$e^{\kappa_L}$ or $e^{-\kappa_L}$ 
respectively to each unfolded  or folded long edge. 
Similar weights  $e^{\pm\kappa_S}$ are associated to 
the two  short edge folding states. 
 
In the plaquette formulation long and short edge bending energies
appear in an asymmetric way.  Long edge bending energy results in an
interaction between each couple of adjacent plaquettes in the loop
formulation.  An energy $-\kappa_L$ has to be assigned to the
network any time a dashed line (or equivalently a full line) crosses
the edge between two plaquettes without bending. In the other case an
energy $\kappa_L$ has to be assigned to any right angle of
dashed lines between neighboring plaquettes (see 
Fig. \ref{fig:due}).
 
Differently, short edge bending energy results in different weigths
for the 32 basic plaquette configurations.  In the reference
orientation state of Fig. \ref{fig:uno}, in each square the two short
edges on the same diagonal always have opposite orientations.  These
two short edges acquire the same sign for each folding relative to
their perpendicular diagonal.  In this case two triangles are folded
and an energy $2\kappa_S$ has to be assigned to the network.
In the loop formulation this is equivalent to have a couple of
parallel lines in one plaquette with the same sign.  Similar
considerations hold for an unfolded edge and for the other couple of
parallel lines in a plaquette.  Therefore, in total, each plaquette
can have an energy equal to $-4\kappa_S$, 0, or $4\kappa_S$ \cite{Di2}.

\subsec{The vertex representation}{sub:vertici}  
\par\noindent 
 From the discussion above it is clear that our model can be thought as 
a vertex model defined on a two--dimensional square lattice ${\mathbb{Z}}^2$. 
A vertex variable $\sigma_x\in\{1,...,32\}$ is associated to each site 
$x$ of the lattice; to each value of $\sigma_x$ corresponds one of the $32$ 
vertices introduced above (a possible representation is shown 
in Fig. \ref{fig:quattro}).
\par
Constraints between neighboring vertices are implied by the sign
conservation law along the closed loops. By using the labelling of
Fig. \ref{fig:quattro} these constraints can be written in a simple
form: we define the two $32\times 32$ matrices
$\mu^h(\sigma,\sigma')=\delta_{u_2(\sigma),u_4(\sigma')}$ and
$\mu^v(\sigma,\sigma')=\delta_{u_3(\sigma),u_1(\sigma')}$ 
where $\sigma,\sigma'\in\{1,...,32\}$ and
$u(\sigma)$ is the vector $u$ associated to the vertex $\sigma$. A
configuration $\bar{\sigma}=\{\sigma_x\}_{{\mathbb{Z}}^2}$ is allowed if
$\mu^h(\sigma_x,\sigma_y)=1$ and $\mu^v(\sigma_x,\sigma_y)=1$
respectively for any horizontal, $\langle x,y\rangle_h$, and vertical,
$\langle x,y\rangle_v$, pair of nearest neighboring sites. Notice that
on each bond there are only $256$ allowed configurations out of the
$32^2=1024$ total ones.  We remark that when we write $\langle
x,y\rangle_{h,v}$ we suppose the sites $x$ and $y$ lexicographically
ordered.
\par
The hamiltonian of the model is in the form 
\begin{equation}
H(\bar{\sigma})=\sum_{x\in{\mathbb{Z}}^2}H^s(\sigma_x) +\sum_{\langle
x,y\rangle_h} H^h(\sigma_x,\sigma_y) +\sum_{\langle x,y\rangle_v}
H^v(\sigma_x,\sigma_y)
\;\;\; ,
\label{eq:hamiltoniana}
\end{equation}
where the two body potential is given by the two $32\times 32$
matrices
$H^v(\sigma',\sigma)=H^h(\sigma,\sigma')
=2\kappa_L\delta_{v(\sigma),1-v(\sigma')}-\kappa_L$
for any $\sigma,\sigma'\in\{1,...,32\}$ 
and the single body potential is given by
the vector $H^s(\sigma)=4\kappa_S$ if $\sigma=1,...,8$,
$H^s(\sigma)=-\kappa_S$ if $\sigma=25,...,32$ and 
$H^s(\sigma)=0$ otherwise.
Finally we can write the partition function of the model:
\begin{equation}
Z(\kappa_S,\kappa_L)=\sum_{\bar{\sigma}}
\left(
\prod_{\langle x,y\rangle_h}\mu^h(\sigma_x,\sigma_y)
\prod_{\langle x,y\rangle_v}\mu^v(\sigma_x,\sigma_y)
\right)
\exp\{-H(\bar{\sigma})\}
\;\;\; ,
\label{eq:partizione}
\end{equation}
where the inverse temperature $\beta$ has been adsorbed in the
hamiltonian. 

\sezione{Ground states}{sect:fondamentali} 
\par\noindent 
The ground states of the network can be conveniently discussed in
terms of loop configurations.  First consider what happens for
$\kappa_S,\kappa_L > 0 $ at very low temperatures.  A large positive
$\kappa_L$ selects states with straight lines so that in a square
lattice with periodic boundary conditions there are the dashed and
full closed loop shown in Fig. \ref{fig:tre}(a).  A positive value of
$\kappa_S$ favors alternate sign values on these lines.  This
corresponds to a completely flat state with an energy per plaquette
given by $E_{1} = - 4 \kappa_S - 2 \kappa_L$.  There are 8
possible degenerate ground states of this kind and each of them can be
realized with two basic plaquette states.
 
In the sector $\kappa_S < 0, \kappa_L > 0 $ elementary
triangles want to be folded along diagonals. A positive value of
$\kappa_L$ still select straight lines. All the dashed lines
have one sign and the same for the full lines.  The network is in a
state where two triangles sharing a long edge are on the same plane
and all the other triangles are above one of those two.  Also in this
case the degeneracy is 8 while one plaquette state is enough for
building the global state which has an energy $E_{2} = 4 
\kappa_S - 2 \kappa_L$ per plaquette and is reported in Fig.\
\ref{fig:tre}(b).
 
Loop--configuration ground states for the remaining sectors 
$\kappa_S < 0, \kappa_L < 0$ and $\kappa_S > 0, 
\kappa_L < 0 $ are respectively shown in Fig. \ref{fig:tre}(c) and
 Fig. \ref{fig:tre}(d). They have an energy equal to $E_{3}= 4 
\kappa_S + 2 \kappa_L$ and $E_{4}= - 4 \kappa_S + 2
 \kappa_L$. They describe a completely folded state with all
 triangle one above the other, and a state with four triangles with a
 common vertex on the same plane and all other triangles
 folded above.  There are needed 2 and 4 different plaquettes for
 building up ground states in sectors 3 and 4 respectively.
 
\sezione{The CVM approximation}{sect:cvm}
\par\noindent 
In this section we describe the CVM approximation we used to study the
phase diagram of model (\ref{eq:partizione}). We do not enter into all
the details related to the CVM technique, we just refer to
\cite{Kik1,An,Morita}.  \par The CVM is based on the minimization of
the free--energy density functional obtained by a truncation of the
cluster (cumulant) expansion of the corresponding functional appearing
in the exact variational formulation of statistical mechanics.  
The existence of different horizontal and vertical interactions in the 
hamiltonian (\ref{eq:hamiltoniana}) and the structure of 
the ground states of the model suggest that one should
consider at least a square (four--vertex) approximation, that is one should
consider a square of four vertices
as the largest cluster in the expansion of the
free--energy functional.  Moreover, in order to reproduce the
structure of the ground states we are forced to partition our lattice
into four square lattices with spacing two (see Fig. \ref{fig:reticolo}). 
We denote by $A$, $B$, $C$
and $D$ these four sublattices and we introduce four
square density matrices: $\rh{ABCD}$, $\rh{BADC}$,
$\rh{CDAB}$ and $\rh{DCBA}$.  We also need four pair
density matrices $\rh{AB}$, $\rh{BA}$, $\rh{CD}$ and $\rh{DC}$
for the four horizontal different bonds, four pair density matrices
$\rh{AC}$, $\rh{CA}$, $\rh{BD}$ and $\rh{DB}$ for the four
different vertical bonds and four single site matrices
$\rh{A}$, $\rh{B}$, $\rh{C}$ and $\rh{D}$.  Finally, we define
the sets ${\cal P}=\{ABCD, BADC, CDAB, DCBA\}$, ${\cal L}^h=\{AB, BA,
CD, DC\}$, ${\cal L}^v=\{AC, CA, BD, DB\}$ and ${\cal S}=\{A,B,C,D\}$,
and we write the free--energy functional per plaquette as follows:
\begin{equation}
\begin{array}{ll}
f(\{\rh{X}:\; X\in{\cal P}\})=&
 \frac{1}{4}\sum_{X\in{\cal L}^h} \Tr{X}\left[\rh{X}H^h\right]   
+\frac{1}{4}\sum_{X\in{\cal L}^v} \Tr{X}\left[\rh{X}H^v\right]    
+\frac{1}{4}\sum_{X\in{\cal S}}   \Tr{X}\left[\rh{X}H^s\right]      
\\
&\\
&
+\frac{1}{4}\left(
 \sum_{X\in{\cal P}}   \Tr{X}\left[\rh{X}\log\rh{X}\right]
-\sum_{X\in{\cal L}^h} \Tr{X}\left[\rh{X}\log\rh{X}\right]
\right.
\\
&\\
&
\left.
-\sum_{X\in{\cal L}^v} \Tr{X}\left[\rh{X}\log\rh{X}\right]
+\sum_{X\in{\cal S}}   \Tr{X}\left[\rh{X}\log\rh{X}\right]
\right)
\\
&\\
&
+\sum_{X\in{\cal P}} \la{X} \left({\mathrm{Tr}}\left[\rh{X}\right]-1\right)
\\
\end{array}
\label{eq:libera}
\end{equation} 
where $\la{X}$, with $X\in{\cal P}$, are four Lagrange multipliers
ensuring that $\rh{X}$ with $X\in{\cal P}$ are correctly normalized
and for any $X\in{\cal P}\cup{\cal L}^h\cup{\cal L}^v\cup{\cal S}$ we
have denoted by $\Tr{X}$ the sum over all the allowed
configurations on the set $X$.  \par Following the recipe of the CVM
one should find the densities $\{\rh{X}:\; X\in{\cal P}\}$ that
minimize the functional (\ref{eq:libera}).  Hence the next step
consists in taking derivatives of the free energy with respect to the
square densities. In order to do this we must take into account
that bond and single site densities can be obtained from square
densities via a partial tracing. Namely,
\begin{equation}
\begin{array}{llll} 
\rh{AB}=\Tr{CD}\rh{ABCD} &   
\rh{BA}=\Tr{DC}\rh{BADC} &   
\rh{CD}=\Tr{AB}\rh{CDAB} &   
\rh{DC}=\Tr{BA}\rh{DCBA} \\
&\\
\rh{AC}=\Tr{BD}\rh{ABCD} &   
\rh{BD}=\Tr{AC}\rh{BADC} &   
\rh{CA}=\Tr{DB}\rh{CDAB} &   
\rh{DB}=\Tr{CA}\rh{DCBA} \\
&\\
\rh{A}=\Tr{B}\rh{AB} &
\rh{B}=\Tr{B}\rh{BA} &
\rh{C}=\Tr{D}\rh{CD} &
\rh{D}=\Tr{D}\rh{DC} \\
\end{array}
\label{eq:parziali}
\end{equation} 
Actually, each bond or single site density can be derived via partial
tracing of different higher order densities. This means that suitable
Lagrange multiplier must be introduced to ensure that different
partial tracings lead to the same result. More precisely, each bond
belongs to two different squares (see Fig. \ref{fig:reticolo}):  
we have to associate a family
of multiplier to each bond to ensure that the same result is obtained 
by tracing over the two plaquettes sharing the bond itself (for instance, 
we need $\Tr{CD}\rh{ABCD}=\Tr{CD}\rh{CDAB}$). We get eight different families
$\{\la{X}:\; X\in{\cal L}^h\cup{\cal L}^v\}$ and each family
contains $256$ different multipliers, one for each allowed bond
configuration.  Moreover, there exist four different bonds sharing the
same single site: to each site we associate three different multiplier
families $\{\la{X,i}:\;X\in{\cal S},\;i=1,2,3\}$, each of them
made of $32$ different multipliers. Hence we have $20$ different
families of multipliers resulting in a total number of $8\times
256+12\times 32=2432$ multipliers.  
\par 
The functional that we have
to minimize is no more the free energy (\ref{eq:libera}), but the one
in which all the Lagrange multipliers are introduced:
\begin{equation}
\begin{array}{ll}
g(\{\rh{X}:\; X\in{\cal P}\})=&
4 f(\{\rh{X}:\; X\in{\cal P}\})
+\Tr{A}\left[\la{A,1}
 \Tr{B}\left[\rh{BA}-\rh{AB}\right]\right]
+\dots\dots\\
&\\
&+\Tr{AB}\left[\la{AB}
  \Tr{CD}\left[\rh{CDAB}-\rh{ABCD}\right]\right]
+\dots\dots\\
\end{array}
\label{eq:funzionale}
\end{equation}
where dots stand for other seven similar bond terms and eleven similar
single site terms.  Now, let us label with $\alpha\in\{1,...,4608\}$
the allowed $4608$ square states. We denote by $\si{X}(\alpha)$
the vertex associated to the site $X\in{\cal S}$ corresponding to
$\alpha$.  To obtain the equilibrium densities we have to set equal
to zero the derivatives of the functional (\ref{eq:funzionale})
taken with respect to $\rh{Y}(\alpha)$ with $Y\in{\cal P}$ and
$\alpha\in\{1,...,4608\}$. In the case $Y=ABCD$ we obtain:
\begin{equation} 
\begin{array}{ll} 
\rh{ABCD}(\alpha)=& 
{\mathrm{const}}\times
\exp\left\{-H^h(\si{A}(\alpha),\si{B}(\alpha)) 
           -H^v(\si{A}(\alpha),\si{C}(\alpha)) 
           -H^s(\si{A}(\alpha))\right.\\
&\\
& +\la{AB}(\si{A}(\alpha),\si{B}(\alpha))
  -\la{CD}(\si{C}(\alpha),\si{D}(\alpha))
  +\la{AC}(\si{A}(\alpha),\si{C}(\alpha))\\ 
&\\
& -\la{BD}(\si{B}(\alpha),\si{D}(\alpha))
  +\la{A,1}(\si{A}(\alpha)) 
  +\la{A,2}(\si{A}(\alpha)) 
  -\la{B,1}(\si{B}(\alpha))\\ 
&\\
&\left.
  -\la{B,3}(\si{B}(\alpha)) 
  -\la{C,2}(\si{C}(\alpha)) 
  +\la{C,3}(\si{C}(\alpha))\right\}\\
&\\
&\times
\rh{AB}(\si{A}(\alpha),\si{B}(\alpha))
\rh{AC}(\si{A}(\alpha),\si{C}(\alpha))
\rh{A}(\si{A}(\alpha))^{-1}\\
\end{array}
\label{eq:iterativa}
\end{equation}
The equation above is actually a set of $4608$ equations. Three
similar sets can be found by considering the cases
$X\in\{BADC,CDAB,DCBA\}$. The complete set of $4\times 4608=18432$
equations, together with the equations for the multipliers, can be
solved by means of the natural iteration method \cite{Kik2,Kik3}.  The
equations for the multipliers can be obtained by taking derivatives of
(\ref{eq:funzionale}) with respect to the multipliers and are of the
form
\begin{equation}
\la{X}(\si{X})=\la{X}(\si{X})+{\mathrm{const}}\log
\frac{\varphi_1(\si{X})}{\varphi_2(\si{X})}
\label{eq:moltiplicatori}
\end{equation}
where $X\in{\cal L}^h\cup{\cal L}^v\cup{\cal S}$, $\si{X}$ is 
an allowed configuration on $X$ and $\varphi_1$ and $\varphi_2$ are 
linear combinations of higher order densities on clusters $Y\supset X$
traced over $Y\setminus X$. 

\sezione{Phase diagram}{sec:diagram} 
\par
\begin{figure}
\begin{center}
\mbox{\epsfig{file=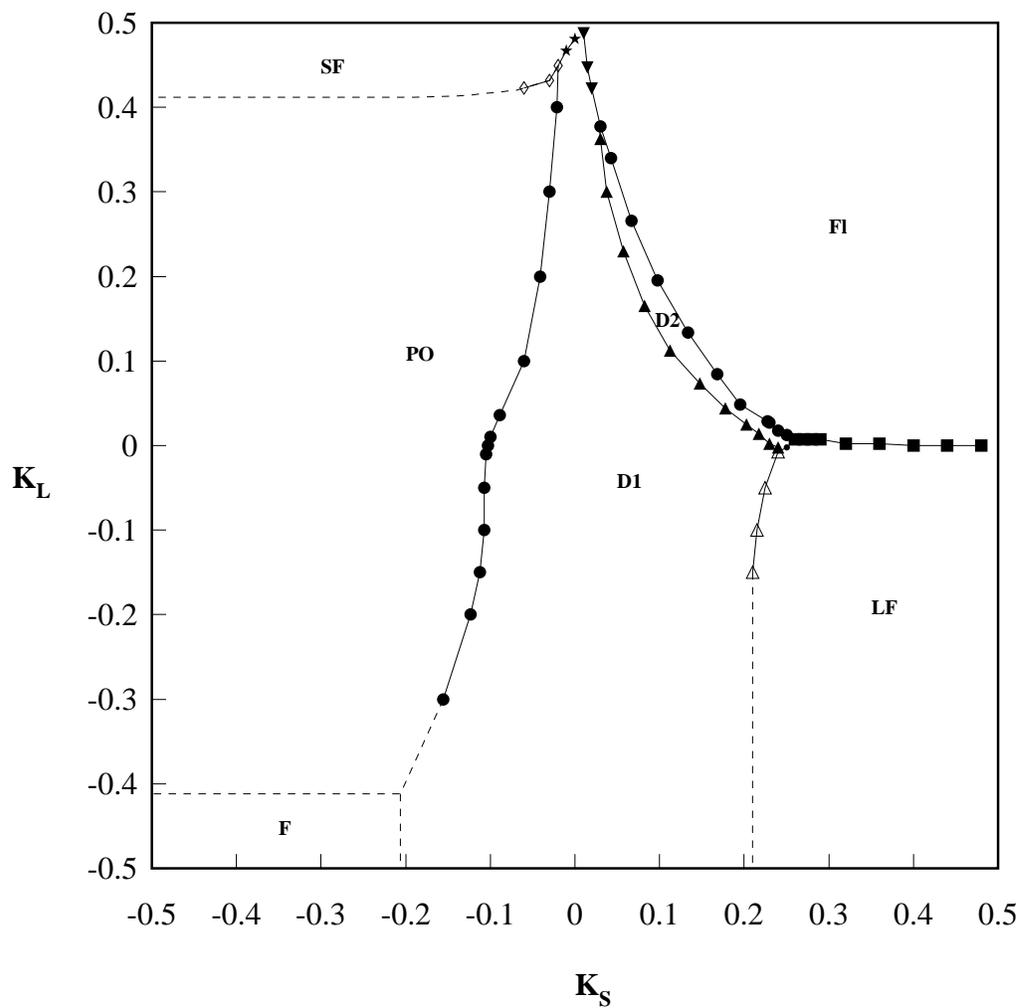, width=15 cm, height=15 cm}}
\end{center}
\vskip 1cm
\caption{The phase diagram: open (full) symbols denote second (first) 
order transition points calculated via CVM. Solid 
lines are eye--guides, dashed lines are suggested by the limits
$\kappa_S$, $\kappa_L\to\infty$.
}
\label{fig:cinque}
\end{figure}

\noindent 
The phase diagram of the model, as predicted by our approximation, has
been reported in Fig. \ref{fig:cinque}, where open symbols denote
second order transitions, and full symbols denote first order
transitions. The transition lines have been drawn using a limited
number of points, since the precise determination of such points is a
very computationally demanding task. In particular, the transitions
between the phase {\bf D1} (Disordered 1) and the phases {\bf F}
(Folded) and {\bf LF} (L--Folded) have been simply sketched since it
was possible to determine their location only with a rough
approximation.

Several phases appear in the diagram. First of all we have four
long--range ordered phases corresponding to the four possible ground
states described in Section \ref{sect:fondamentali}. Phase {\bf Fl} (Flat) is
stable for large and positive $\kappa_S$ and $\kappa_L$ and
represents a flat phase. Phase {\bf SF} (S--Folded) is stable for large enough
absolute values of $\kappa_S < 0$ and $\kappa_L > 0$. It
represents a partially folded phase in which folding occurs mainly
along short edges. Phase {\bf F} is stable for large enough absolute
values of $\kappa_S < 0$ and $\kappa_L < 0$. It represents a
completely folded phase. Phase {\bf LF} is stable for large enough
absolute values of $\kappa_S > 0$ and $\kappa_L < 0$. It
represents a partially folded phase in which folding occurs mainly at
long edges. Like the corresponding ground states, all these phases
have degeneracy 8.

On the high temperature side of the ordered flat phase {\bf Fl} we have
a small slice of the disordered phase {\bf D2} (Disordered 2).  
In the central part of
the phase diagram, we have the disordered phase {\bf D1}, which has
larger entropy and larger energy than {\bf D2}. It is noteworthy that
the entropy of this phase at the infinite temperature point
$\kappa_S = \kappa_L = 0$ is $S_\infty \simeq 0.9204$ in our
approximation, while the estimate in \cite{Di1}, obtained by 
transfer matrix methods, 
corresponds, with our definitions, to $S_\infty \simeq 0.9196$.

Between phases {\bf SF} and {\bf F} we have the partially ordered phase
{\bf PO} (Partially Ordered). 
For large negative $\kappa_S$ the transitions between
this phase and phases {\bf SF} and {\bf F} tend asymptotically to
$\vert \kappa_L \vert = \frac{1}{2} \ln \frac{5 +
\sqrt{17}}{4} \simeq 0.412$, which is exactly the estimate for the
critical coupling of the square lattice Ising model in the present
approximation \cite{Kik1}. This is a consequence of a property of the
model, which reduces to a square lattice Ising model in the limit
$\kappa_S \to - \infty$. In the loop gas formulation,
$\kappa_S \to - \infty$ implies that all loops in the same set
(that is, all dashed loops, or all full loops) must have the same
sign. Looking at Fig.\ \ref{fig:quattro} one can verify that to
satisfy this condition the state of the system must be a mixture of
only 2 out of the 32 plaquette states. For this pair of allowed states
there are four different possibilities (and thus the phase {\bf PO}
has degeneracy 4): (1,5) (corresponding to $+$ signs on both loop
sets), (2,6) ($+$ on full loops and $-$ on dashed loops), (3,7) ($-$
on both loop sets) and (4,8) ($-$ on full loops and $+$ on dashed
loops). Hence we have the breaking of the loop sign inversion
symmetry, which is restored only at the transition to the phase {\bf
5}. Given a pair of allowed states, the hamiltonian reduces to an
ordinary Ising hamiltonian, with equal states on adjacent plaquettes
giving a contribution $-\kappa_L$ to the total energy, and
different states giving $+\kappa_L$. It is therefore an exact
result that, in the limit $\kappa_S \to - \infty$, there must
exist the three phases {\bf SF}, {\bf F} and {\bf PO} (corresponding
respectively to the ferromagnetic, antiferromagnetic and disordered
phases of the Ising model), separated by second order phase
transitions at $\vert\kappa_{Lc} \vert =
\frac{1}{2} \ln \left(1 + \sqrt{2} \right) \simeq
0.441$.

A similar situation occurs in the limit $\kappa_L \to - \infty$.
Looking at Fig.\ \ref{fig:tre} one sees that in this limit both the
full and the dashed lines form the smallest possible square loops. It
follows that the model reduces to two decoupled Ising models, where
the Ising variables are the loop signs and two loops (both full, or
both dashed) having parallel edges in the same plaquette interact with
an energy $\pm 2 \kappa_S$. Therefore the phases {\bf F} and {\bf LF}
(corresponding respectively to the antiferromagnetic and ferromagnetic
phases of the limiting Ising model) will undergo second order phase
transitions towards the disordered phase {\bf D1} at $\vert\kappa_{Sc}
\vert = \frac{1}{4} \ln \left(1 + \sqrt{2} \right) \simeq 0.220$. The
estimate for this value in our approximation is of course $\vert
\kappa_L \vert = \frac{1}{4} \ln \frac{5 + \sqrt{17}}{4} \simeq
0.206$.

Comparing our phase diagram with that by Di Francesco \cite{Di2} one
can see several striking differences. First of all, among our ordered
phases, only the completely flat and completely folded ones were
reported in \cite{Di2}, and no low temperature transition was found
among them. In addition, no intermediate phase (like our phases 
{\bf PO} and {\bf D2}) was found between the ordered and the disordered
phases. 
We note that the CVM results 
are fully confirmed by the ground states analysis and the limits
$\kappa_S \to - \infty$ and $\kappa_L \to - \infty$.
\par
In conclusions, we have shown that the CVM
techniques can be adapted to study very complex 
vertex models as the one considered in this paper 
representing the folding configurations of a square--diagonal lattice.
As in the case of the folding of the triangular lattice, 
a first--order crumpling transition between the flat phase and a
disordered phase has been found. The extension of this analysis 
to the $d$--dimensional folding problem appears not easy from a
numerical point of view.

\bigskip 
\par\noindent 
{\textbf{Acknowledgements}} 
\par\noindent 
G.G. acknowledges  
support by MURST (COFIN97). E.C. 
whishes to express his thanks to the European  
network ``Stochastic Analysis and its Applications" ERB--FMRX--CT96--0075
for financial support.

\setlength{\unitlength}{3pt} 
\newpage 
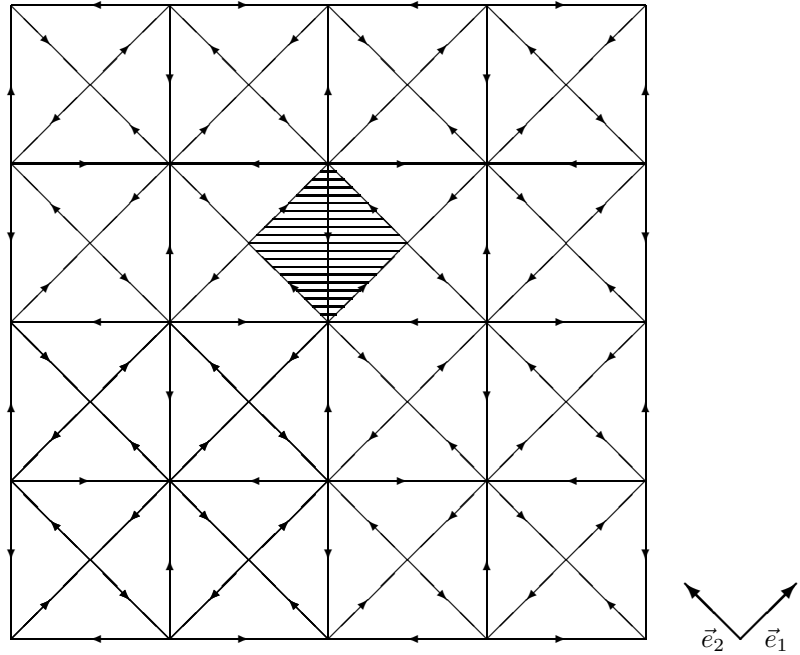
\begin{figure} 
\begin{center} 
\begin{picture}(180,180)(-35,-100) 
\thinlines 
 
\multiput(0,10)(40,0){3}{\line(0,-1){10}} 
\multiput(0,20)(40,0){3}{\vector(0,-1){10}} 
\multiput(20,30)(40,0){2}{\line(0,-1){10}} 
\multiput(20,40)(40,0){2}{\vector(0,-1){10}} 
\multiput(0,50)(40,0){3}{\line(0,-1){10}} 
\multiput(0,60)(40,0){3}{\vector(0,-1){10}} 
\multiput(20,70)(40,0){2}{\line(0,-1){10}} 
\multiput(20,80)(40,0){2}{\vector(0,-1){10}} 
\multiput(20,0)(40,0){2}{\vector(0,1){10}} 
\multiput(20,10)(40,0){2}{\line(0,1){10}} 
\multiput(0,20)(40,0){3}{\vector(0,1){10}} 
\multiput(0,30)(40,0){3}{\line(0,1){10}} 
\multiput(20,40)(40,0){2}{\vector(0,1){10}} 
\multiput(20,50)(40,0){2}{\line(0,1){10}} 
\multiput(0,60)(40,0){3}{\vector(0,1){10}} 
\multiput(0,70)(40,0){3}{\line(0,1){10}} 
\multiput(10,0)(0,40){3}{\line(-1,0){10}} 
\multiput(20,0)(0,40){3}{\vector(-1,0){10}} 
\multiput(30,20)(0,40){2}{\line(-1,0){10}} 
\multiput(40,20)(0,40){2}{\vector(-1,0){10}} 
\multiput(50,0)(0,40){3}{\line(-1,0){10}} 
\multiput(60,0)(0,40){3}{\vector(-1,0){10}} 
\multiput(70,20)(0,40){2}{\line(-1,0){10}} 
\multiput(80,20)(0,40){2}{\vector(-1,0){10}} 
\multiput(0,20)(0,40){2}{\vector(1,0){10}} 
\multiput(10,20)(0,40){2}{\line(1,0){10}} 
\multiput(20,0)(0,40){3}{\vector(1,0){10}} 
\multiput(30,0)(0,40){3}{\line(1,0){10}} 
\multiput(40,20)(0,40){2}{\vector(1,0){10}} 
\multiput(50,20)(0,40){2}{\line(1,0){10}} 
\multiput(60,0)(0,40){3}{\vector(1,0){10}} 
\multiput(70,0)(0,40){3}{\line(1,0){10}} 
\multiput(0,0)(40,0){2}{\vector(1,1){5}} 
\multiput(5,5)(40,0){2}{\line(1,1){5}} 
\multiput(0,0)(0,40){2}{\vector(1,1){5}} 
\multiput(5,5)(0,40){2}{\line(1,1){5}} 
\multiput(0,0)(40,40){2}{\vector(1,1){5}} 
\multiput(5,5)(40,40){2}{\line(1,1){5}} 
 
\multiput(30,10)(40,0){2}{\vector(1,1){5}} 
\multiput(35,15)(40,0){2}{\line(1,1){5}} 
\multiput(30,10)(0,40){2}{\vector(1,1){5}} 
\multiput(35,15)(0,40){2}{\line(1,1){5}} 
\multiput(30,10)(40,40){2}{\vector(1,1){5}} 
\multiput(35,15)(40,40){2}{\line(1,1){5}} 
 
\multiput(10,30)(40,0){2}{\vector(1,1){5}} 
\multiput(15,35)(40,0){2}{\line(1,1){5}} 
\multiput(10,30)(0,40){2}{\vector(1,1){5}} 
\multiput(15,35)(0,40){2}{\line(1,1){5}} 
\multiput(10,30)(40,40){2}{\vector(1,1){5}} 
\multiput(15,35)(40,40){2}{\line(1,1){5}} 
 
\multiput(20,20)(40,0){2}{\vector(1,1){5}} 
\multiput(25,25)(40,0){2}{\line(1,1){5}} 
\multiput(20,20)(0,40){2}{\vector(1,1){5}} 
\multiput(25,25)(0,40){2}{\line(1,1){5}} 
\multiput(20,20)(40,40){2}{\vector(1,1){5}} 
\multiput(25,25)(40,40){2}{\line(1,1){5}} 
 
\multiput(10,10)(40,0){2}{\vector(-1,1){5}} 
\multiput(5,15)(40,0){2}{\line(-1,1){5}} 
\multiput(10,10)(0,40){2}{\vector(-1,1){5}} 
\multiput(5,15)(0,40){2}{\line(-1,1){5}} 
\multiput(10,10)(40,40){2}{\vector(-1,1){5}} 
\multiput(5,15)(40,40){2}{\line(-1,1){5}} 
 
\multiput(40,0)(40,0){2}{\vector(-1,1){5}} 
\multiput(35,5)(40,0){2}{\line(-1,1){5}} 
\multiput(40,0)(0,40){2}{\vector(-1,1){5}} 
\multiput(35,5)(0,40){2}{\line(-1,1){5}} 
\multiput(40,0)(40,40){2}{\vector(-1,1){5}} 
\multiput(35,5)(40,40){2}{\line(-1,1){5}} 
 
\multiput(20,20)(40,0){2}{\vector(-1,1){5}} 
\multiput(15,25)(40,0){2}{\line(-1,1){5}} 
\multiput(20,20)(0,40){2}{\vector(-1,1){5}} 
\multiput(15,25)(0,40){2}{\line(-1,1){5}} 
\multiput(20,20)(40,40){2}{\vector(-1,1){5}} 
\multiput(15,25)(40,40){2}{\line(-1,1){5}} 
 
\multiput(30,30)(40,0){2}{\vector(-1,1){5}} 
\multiput(25,35)(40,0){2}{\line(-1,1){5}} 
\multiput(30,30)(0,40){2}{\vector(-1,1){5}} 
\multiput(25,35)(0,40){2}{\line(-1,1){5}} 
\multiput(30,30)(40,40){2}{\vector(-1,1){5}} 
\multiput(25,35)(40,40){2}{\line(-1,1){5}} 
 
\multiput(10,10)(40,0){2}{\vector(1,-1){5}} 
\multiput(15,5)(40,0){2}{\line(1,-1){5}} 
\multiput(10,10)(0,40){2}{\vector(1,-1){5}} 
\multiput(15,5)(0,40){2}{\line(1,-1){5}} 
\multiput(10,10)(40,40){2}{\vector(1,-1){5}} 
\multiput(15,5)(40,40){2}{\line(1,-1){5}} 
 
\multiput(20,20)(40,0){2}{\vector(1,-1){5}} 
\multiput(25,15)(40,0){2}{\line(1,-1){5}} 
\multiput(20,20)(0,40){2}{\vector(1,-1){5}} 
\multiput(25,15)(0,40){2}{\line(1,-1){5}} 
\multiput(20,20)(40,40){2}{\vector(1,-1){5}} 
\multiput(25,15)(40,40){2}{\line(1,-1){5}} 
 
\multiput(0,40)(40,0){2}{\vector(1,-1){5}} 
\multiput(5,35)(40,0){2}{\line(1,-1){5}} 
\multiput(0,40)(0,40){2}{\vector(1,-1){5}} 
\multiput(5,35)(0,40){2}{\line(1,-1){5}} 
\multiput(0,40)(40,40){2}{\vector(1,-1){5}} 
\multiput(5,35)(40,40){2}{\line(1,-1){5}} 
 
\multiput(30,30)(40,0){2}{\vector(1,-1){5}} 
\multiput(35,25)(40,0){2}{\line(1,-1){5}} 
\multiput(30,30)(0,40){2}{\vector(1,-1){5}} 
\multiput(35,25)(0,40){2}{\line(1,-1){5}} 
\multiput(30,30)(40,40){2}{\vector(1,-1){5}} 
\multiput(35,25)(40,40){2}{\line(1,-1){5}} 
 
\multiput(20,20)(40,0){2}{\vector(-1,-1){5}} 
\multiput(15,15)(40,0){2}{\line(-1,-1){5}} 
\multiput(20,20)(0,40){2}{\vector(-1,-1){5}} 
\multiput(15,15)(0,40){2}{\line(-1,-1){5}} 
\multiput(20,20)(40,40){2}{\vector(-1,-1){5}} 
\multiput(15,15)(40,40){2}{\line(-1,-1){5}} 
 
\multiput(30,10)(40,0){2}{\vector(-1,-1){5}} 
\multiput(25,5)(40,0){2}{\line(-1,-1){5}} 
\multiput(30,10)(0,40){2}{\vector(-1,-1){5}} 
\multiput(25,5)(0,40){2}{\line(-1,-1){5}} 
\multiput(30,10)(40,40){2}{\vector(-1,-1){5}} 
\multiput(25,5)(40,40){2}{\line(-1,-1){5}} 
 
\multiput(10,30)(40,0){2}{\vector(-1,-1){5}} 
\multiput(5,25)(40,0){2}{\line(-1,-1){5}} 
\multiput(10,30)(0,40){2}{\vector(-1,-1){5}} 
\multiput(5,25)(0,40){2}{\line(-1,-1){5}} 
\multiput(10,30)(40,40){2}{\vector(-1,-1){5}} 
\multiput(5,25)(40,40){2}{\line(-1,-1){5}} 
 
\multiput(40,40)(40,0){2}{\vector(-1,-1){5}} 
\multiput(35,35)(40,0){2}{\line(-1,-1){5}} 
\multiput(40,40)(0,40){2}{\vector(-1,-1){5}} 
\multiput(35,35)(0,40){2}{\line(-1,-1){5}} 
\multiput(40,40)(40,40){2}{\vector(-1,-1){5}} 
\multiput(35,35)(40,40){2}{\line(-1,-1){5}} 
 
\thicklines 
\put(92,0){\vector(1,1){7}} 
\put(92,0){\vector(-1,1){7}} 
\put(95,-1){$\vec{e}_1$} 
\put(87,-1){$\vec{e}_2$} 

\thinlines
\put(39,41){\line(1,0){2}}
\put(38,42){\line(1,0){4}}
\put(37,43){\line(1,0){6}}
\put(36,44){\line(1,0){8}}
\put(35,45){\line(1,0){10}}
\put(34,46){\line(1,0){12}}
\put(33,47){\line(1,0){14}}
\put(32,48){\line(1,0){16}}
\put(31,49){\line(1,0){18}}
\put(30,50){\line(1,0){20}}
\put(39,59){\line(1,0){2}}
\put(38,58){\line(1,0){4}}
\put(37,57){\line(1,0){6}}
\put(36,56){\line(1,0){8}}
\put(35,55){\line(1,0){10}}
\put(34,54){\line(1,0){12}}
\put(33,53){\line(1,0){14}}
\put(32,52){\line(1,0){16}}
\put(31,51){\line(1,0){18}}

\end{picture} 
\end{center} 
\vskip -0.5 cm 
\caption{The square--diagonal lattice.} 
\label{fig:uno} 
\end{figure}

\setlength{\unitlength}{2pt} 
\newpage 
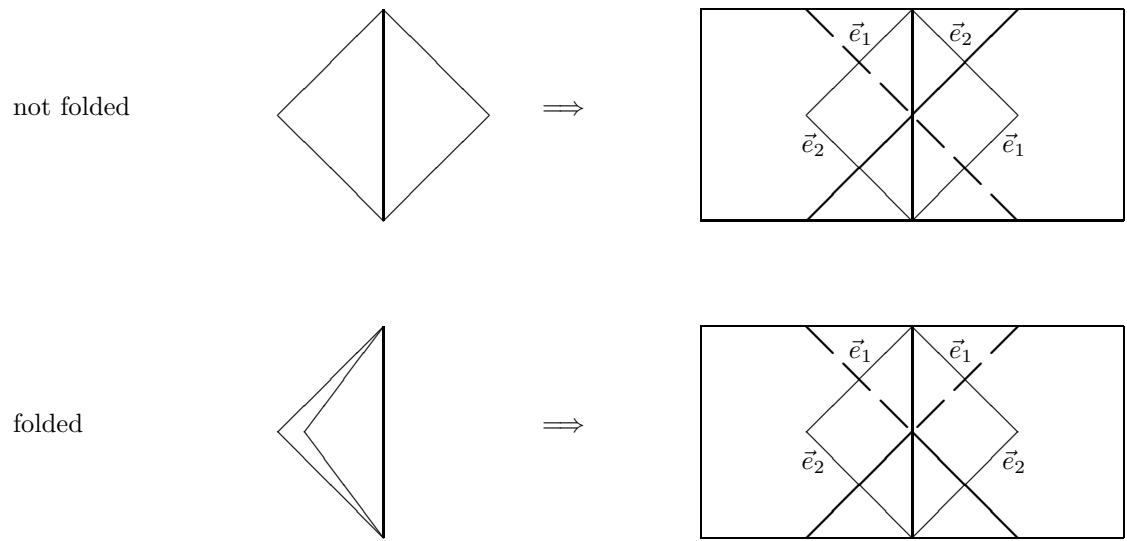
\begin{figure} 
\begin{center} 
\begin{picture}(180,180)(0,-100) 
\thinlines 
 
\put(-20,100){not folded} 
\put(30,100){\line(1,1){20}} 
\put(30,100){\line(1,-1){20}} 
\put(50,80){\line(0,1){40}} 
\put(70,100){\line(-1,1){20}} 
\put(70,100){\line(-1,-1){20}} 
\put(80,100){$\Longrightarrow$} 
\put(110,80){\line(0,1){40}} 
\put(110,120){\line(1,0){80}} 
\put(110,80){\line(1,0){80}} 
\put(190,80){\line(0,1){40}} 
 
\put(130,100){\line(1,1){20}} 
\put(130,100){\line(1,-1){20}} 
\put(150,80){\line(0,1){40}} 
\put(170,100){\line(-1,1){20}} 
\put(170,100){\line(-1,-1){20}} 
 
\thicklines 
\put(130,80){\line(1,1){40}} 
\multiput(130,120)(7,-7){6}{\line(1,-1){5}} 

\put(138,114){$\vec{e}_1$}
\put(167,93){$\vec{e}_1$}
\put(157,114){$\vec{e}_2$}
\put(129,93){$\vec{e}_2$}
 
\thinlines 
\put(-20,40){folded} 
\put(30,40){\line(1,1){20}} 
\put(30,40){\line(1,-1){20}} 
\put(50,20){\line(0,1){40}} 
\put(35,40){\line(3,4){15}} 
\put(35,40){\line(3,-4){15}} 
\put(80,40){$\Longrightarrow$} 
\put(110,20){\line(0,1){40}} 
\put(110,60){\line(1,0){80}} 
\put(110,20){\line(1,0){80}} 
\put(190,20){\line(0,1){40}} 
 
\put(130,40){\line(1,1){20}} 
\put(130,40){\line(1,-1){20}} 
\put(150,20){\line(0,1){40}} 
\put(170,40){\line(-1,1){20}} 
\put(170,40){\line(-1,-1){20}} 
 
\thicklines 
\put(130,20){\line(1,1){20}} 
\put(170,20){\line(-1,1){20}} 
\multiput(130,60)(7,-7){3}{\line(1,-1){5}} 
\multiput(170,60)(-7,-7){3}{\line(-1,-1){5}} 
 
\put(138,54){$\vec{e}_1$}
\put(167,33){$\vec{e}_2$}
\put(157,54){$\vec{e}_1$}
\put(129,33){$\vec{e}_2$}

\end{picture} 
\end{center} 
\vskip -0.5 cm 
\caption{The mapping onto the vertex model.} 
\label{fig:due} 
\end{figure} 

\newpage 
\setlength{\unitlength}{1pt} 
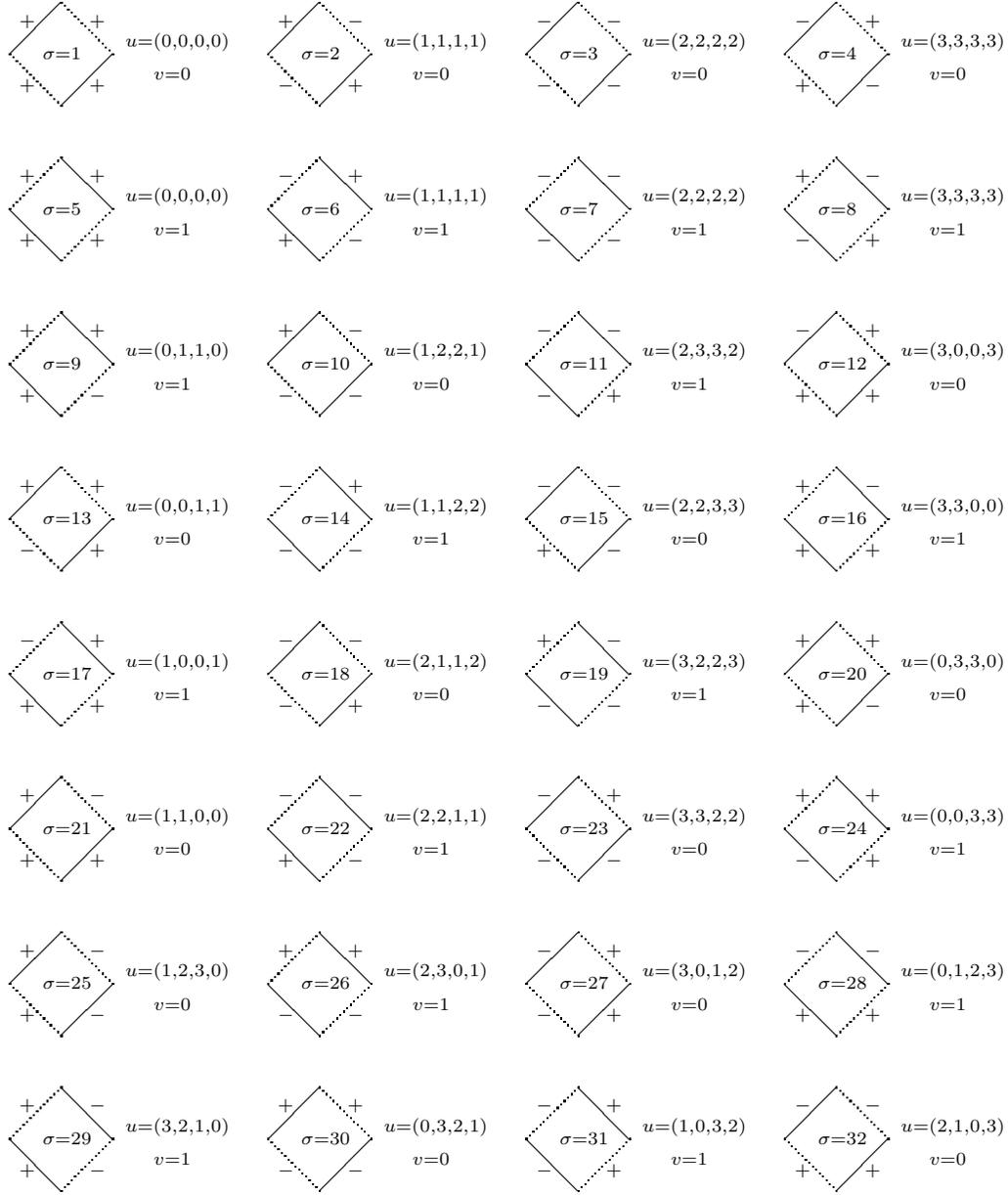
\begin{figure} 
\begin{center} 
\begin{picture}(300,300)(50,-150) 
\thinlines 

\put(0,280){\line(1,1){20}}
\put(40,280){\line(-1,-1){20}}
\qbezier[15](0,280)(10,270)(20,260)
\qbezier[15](20,300)(30,290)(40,280)
\put(4,291){$\scriptstyle{+}$}
\put(4,266){$\scriptstyle{+}$}
\put(31,266){$\scriptstyle{+}$}
\put(31,291){$\scriptstyle{+}$}
\put(13,278){$\scriptstyle{\sigma=1}$}
\put(45,283){$\scriptstyle{u=(0,0,0,0)}$}
\put(56,270){$\scriptstyle{v=0}$}
 
\put(100,280){\line(1,1){20}}
\put(140,280){\line(-1,-1){20}}
\qbezier[15](100,280)(110,270)(120,260)
\qbezier[15](120,300)(130,290)(140,280)
\put(104,291){$\scriptstyle{+}$}
\put(104,266){$\scriptstyle{-}$}
\put(131,266){$\scriptstyle{+}$}
\put(131,291){$\scriptstyle{-}$}
\put(113,278){$\scriptstyle{\sigma=2}$}
\put(145,283){$\scriptstyle{u=(1,1,1,1)}$}
\put(156,270){$\scriptstyle{v=0}$}

\put(200,280){\line(1,1){20}}
\put(240,280){\line(-1,-1){20}}
\qbezier[15](200,280)(210,270)(220,260)
\qbezier[15](220,300)(230,290)(240,280)
\put(204,291){$\scriptstyle{-}$}
\put(204,266){$\scriptstyle{-}$}
\put(231,266){$\scriptstyle{-}$}
\put(231,291){$\scriptstyle{-}$}
\put(213,278){$\scriptstyle{\sigma=3}$}
\put(245,283){$\scriptstyle{u=(2,2,2,2)}$}
\put(256,270){$\scriptstyle{v=0}$}

\put(300,280){\line(1,1){20}}
\put(340,280){\line(-1,-1){20}}
\qbezier[15](300,280)(310,270)(320,260)
\qbezier[15](320,300)(330,290)(340,280)
\put(304,291){$\scriptstyle{-}$}
\put(304,266){$\scriptstyle{+}$}
\put(331,266){$\scriptstyle{-}$}
\put(331,291){$\scriptstyle{+}$}
\put(313,278){$\scriptstyle{\sigma=4}$}
\put(345,283){$\scriptstyle{u=(3,3,3,3)}$}
\put(356,270){$\scriptstyle{v=0}$}

\put(0,220){\line(1,-1){20}}
\put(40,220){\line(-1,1){20}}
\qbezier[15](0,220)(10,230)(20,240)
\qbezier[15](20,200)(30,210)(40,220)
\put(4,231){$\scriptstyle{+}$}
\put(4,206){$\scriptstyle{+}$}
\put(31,206){$\scriptstyle{+}$}
\put(31,231){$\scriptstyle{+}$}
\put(13,218){$\scriptstyle{\sigma=5}$}
\put(45,223){$\scriptstyle{u=(0,0,0,0)}$}
\put(56,210){$\scriptstyle{v=1}$}
 
\put(100,220){\line(1,-1){20}}
\put(140,220){\line(-1,1){20}}
\qbezier[15](100,220)(110,230)(120,240)
\qbezier[15](120,200)(130,210)(140,220)
\put(104,231){$\scriptstyle{-}$}
\put(104,206){$\scriptstyle{+}$}
\put(131,206){$\scriptstyle{-}$}
\put(131,231){$\scriptstyle{+}$}
\put(113,218){$\scriptstyle{\sigma=6}$}
\put(145,223){$\scriptstyle{u=(1,1,1,1)}$}
\put(156,210){$\scriptstyle{v=1}$}

\put(200,220){\line(1,-1){20}}
\put(240,220){\line(-1,1){20}}
\qbezier[15](200,220)(210,230)(220,240)
\qbezier[15](220,200)(230,210)(240,220)
\put(204,231){$\scriptstyle{-}$}
\put(204,206){$\scriptstyle{-}$}
\put(231,206){$\scriptstyle{-}$}
\put(231,231){$\scriptstyle{-}$}
\put(213,218){$\scriptstyle{\sigma=7}$}
\put(245,223){$\scriptstyle{u=(2,2,2,2)}$}
\put(256,210){$\scriptstyle{v=1}$}

\put(300,220){\line(1,-1){20}}
\put(340,220){\line(-1,1){20}}
\qbezier[15](300,220)(310,230)(320,240)
\qbezier[15](320,200)(330,210)(340,220)
\put(304,231){$\scriptstyle{+}$}
\put(304,206){$\scriptstyle{-}$}
\put(331,206){$\scriptstyle{+}$}
\put(331,231){$\scriptstyle{-}$}
\put(313,218){$\scriptstyle{\sigma=8}$}
\put(345,223){$\scriptstyle{u=(3,3,3,3)}$}
\put(356,210){$\scriptstyle{v=1}$}

\put(0,160){\line(1,-1){20}}
\put(40,160){\line(-1,1){20}}
\qbezier[15](0,160)(10,170)(20,180)
\qbezier[15](20,140)(30,150)(40,160)
\put(4,171){$\scriptstyle{+}$}
\put(4,146){$\scriptstyle{+}$}
\put(31,146){$\scriptstyle{-}$}
\put(31,171){$\scriptstyle{+}$}
\put(13,158){$\scriptstyle{\sigma=9}$}
\put(45,163){$\scriptstyle{u=(0,1,1,0)}$}
\put(56,150){$\scriptstyle{v=1}$}

\put(100,160){\line(1,1){20}}
\put(140,160){\line(-1,-1){20}}
\qbezier[15](100,160)(110,150)(120,140)
\qbezier[15](120,180)(130,170)(140,160)
\put(104,171){$\scriptstyle{+}$}
\put(104,146){$\scriptstyle{-}$}
\put(131,146){$\scriptstyle{-}$}
\put(131,171){$\scriptstyle{-}$}
\put(113,158){$\scriptstyle{\sigma=10}$}
\put(145,163){$\scriptstyle{u=(1,2,2,1)}$}
\put(156,150){$\scriptstyle{v=0}$}

\put(200,160){\line(1,-1){20}}
\put(240,160){\line(-1,1){20}}
\qbezier[15](200,160)(210,170)(220,180)
\qbezier[15](220,140)(230,150)(240,160)
\put(204,171){$\scriptstyle{-}$}
\put(204,146){$\scriptstyle{-}$}
\put(231,146){$\scriptstyle{+}$}
\put(231,171){$\scriptstyle{-}$}
\put(213,158){$\scriptstyle{\sigma=11}$}
\put(245,163){$\scriptstyle{u=(2,3,3,2)}$}
\put(256,150){$\scriptstyle{v=1}$}

\put(300,160){\line(1,1){20}}
\put(340,160){\line(-1,-1){20}}
\qbezier[15](300,160)(310,150)(320,140)
\qbezier[15](320,180)(330,170)(340,160)
\put(304,171){$\scriptstyle{-}$}
\put(304,146){$\scriptstyle{+}$}
\put(331,146){$\scriptstyle{+}$}
\put(331,171){$\scriptstyle{+}$}
\put(313,158){$\scriptstyle{\sigma=12}$}
\put(345,163){$\scriptstyle{u=(3,0,0,3)}$}
\put(356,150){$\scriptstyle{v=0}$}

\put(0,100){\line(1,1){20}}
\put(40,100){\line(-1,-1){20}}
\qbezier[15](0,100)(10,90)(20,80)
\qbezier[15](20,120)(30,110)(40,100)
\put(4,111){$\scriptstyle{+}$}
\put(4,86){$\scriptstyle{-}$}
\put(31,86){$\scriptstyle{+}$}
\put(31,111){$\scriptstyle{+}$}
\put(13,98){$\scriptstyle{\sigma=13}$}
\put(45,103){$\scriptstyle{u=(0,0,1,1)}$}
\put(56,90){$\scriptstyle{v=0}$}

\put(100,100){\line(1,-1){20}}
\put(140,100){\line(-1,1){20}}
\qbezier[15](100,100)(110,110)(120,120)
\qbezier[15](120,80)(130,90)(140,100)
\put(104,111){$\scriptstyle{-}$}
\put(104,86){$\scriptstyle{-}$}
\put(131,86){$\scriptstyle{-}$}
\put(131,111){$\scriptstyle{+}$}
\put(113,98){$\scriptstyle{\sigma=14}$}
\put(145,103){$\scriptstyle{u=(1,1,2,2)}$}
\put(156,90){$\scriptstyle{v=1}$}

\put(200,100){\line(1,1){20}}
\put(240,100){\line(-1,-1){20}}
\qbezier[15](200,100)(210,90)(220,80)
\qbezier[15](220,120)(230,110)(240,100)
\put(204,111){$\scriptstyle{-}$}
\put(204,86){$\scriptstyle{+}$}
\put(231,86){$\scriptstyle{-}$}
\put(231,111){$\scriptstyle{-}$}
\put(213,98){$\scriptstyle{\sigma=15}$}
\put(245,103){$\scriptstyle{u=(2,2,3,3)}$}
\put(256,90){$\scriptstyle{v=0}$}

\put(300,100){\line(1,-1){20}}
\put(340,100){\line(-1,1){20}}
\qbezier[15](300,100)(310,110)(320,120)
\qbezier[15](320,80)(330,90)(340,100)
\put(304,111){$\scriptstyle{+}$}
\put(304,86){$\scriptstyle{+}$}
\put(331,86){$\scriptstyle{+}$}
\put(331,111){$\scriptstyle{-}$}
\put(313,98){$\scriptstyle{\sigma=16}$}
\put(345,103){$\scriptstyle{u=(3,3,0,0)}$}
\put(356,90){$\scriptstyle{v=1}$}

\put(0,40){\line(1,-1){20}}
\put(40,40){\line(-1,1){20}}
\qbezier[15](0,40)(10,50)(20,60)
\qbezier[15](20,20)(30,30)(40,40)
\put(4,51){$\scriptstyle{-}$}
\put(4,26){$\scriptstyle{+}$}
\put(31,26){$\scriptstyle{+}$}
\put(31,51){$\scriptstyle{+}$}
\put(13,38){$\scriptstyle{\sigma=17}$}
\put(45,43){$\scriptstyle{u=(1,0,0,1)}$}
\put(56,30){$\scriptstyle{v=1}$}

\put(100,40){\line(1,1){20}}
\put(140,40){\line(-1,-1){20}}
\qbezier[15](100,40)(110,30)(120,20)
\qbezier[15](120,60)(130,50)(140,40)
\put(104,51){$\scriptstyle{-}$}
\put(104,26){$\scriptstyle{-}$}
\put(131,26){$\scriptstyle{+}$}
\put(131,51){$\scriptstyle{-}$}
\put(113,38){$\scriptstyle{\sigma=18}$}
\put(145,43){$\scriptstyle{u=(2,1,1,2)}$}
\put(156,30){$\scriptstyle{v=0}$}

\put(200,40){\line(1,-1){20}}
\put(240,40){\line(-1,1){20}}
\qbezier[15](200,40)(210,50)(220,60)
\qbezier[15](220,20)(230,30)(240,40)
\put(204,51){$\scriptstyle{+}$}
\put(204,26){$\scriptstyle{-}$}
\put(231,26){$\scriptstyle{-}$}
\put(231,51){$\scriptstyle{-}$}
\put(213,38){$\scriptstyle{\sigma=19}$}
\put(245,43){$\scriptstyle{u=(3,2,2,3)}$}
\put(256,30){$\scriptstyle{v=1}$}

\put(300,40){\line(1,1){20}}
\put(340,40){\line(-1,-1){20}}
\qbezier[15](300,40)(310,30)(320,20)
\qbezier[15](320,60)(330,50)(340,40)
\put(304,51){$\scriptstyle{+}$}
\put(304,26){$\scriptstyle{+}$}
\put(331,26){$\scriptstyle{-}$}
\put(331,51){$\scriptstyle{+}$}
\put(313,38){$\scriptstyle{\sigma=20}$}
\put(345,43){$\scriptstyle{u=(0,3,3,0)}$}
\put(356,30){$\scriptstyle{v=0}$}

\put(0,-20){\line(1,1){20}}
\put(40,-20){\line(-1,-1){20}}
\qbezier[15](0,-20)(10,-30)(20,-40)
\qbezier[15](20,0)(30,-10)(40,-20)
\put(4,-9){$\scriptstyle{+}$}
\put(4,-34){$\scriptstyle{+}$}
\put(31,-34){$\scriptstyle{+}$}
\put(31,-9){$\scriptstyle{-}$}
\put(13,-22){$\scriptstyle{\sigma=21}$}
\put(45,-17){$\scriptstyle{u=(1,1,0,0)}$}
\put(56,-30){$\scriptstyle{v=0}$}

\put(100,-20){\line(1,-1){20}}
\put(140,-20){\line(-1,1){20}}
\qbezier[15](100,-20)(110,-10)(120,0)
\qbezier[15](120,-40)(130,-30)(140,-20)
\put(104,-9){$\scriptstyle{-}$}
\put(104,-34){$\scriptstyle{+}$}
\put(131,-34){$\scriptstyle{-}$}
\put(131,-9){$\scriptstyle{-}$}
\put(113,-22){$\scriptstyle{\sigma=22}$}
\put(145,-17){$\scriptstyle{u=(2,2,1,1)}$}
\put(156,-30){$\scriptstyle{v=1}$}

\put(200,-20){\line(1,1){20}}
\put(240,-20){\line(-1,-1){20}}
\qbezier[15](200,-20)(210,-30)(220,-40)
\qbezier[15](220,0)(230,-10)(240,-20)
\put(204,-9){$\scriptstyle{-}$}
\put(204,-34){$\scriptstyle{-}$}
\put(231,-34){$\scriptstyle{-}$}
\put(231,-9){$\scriptstyle{+}$}
\put(213,-22){$\scriptstyle{\sigma=23}$}
\put(245,-17){$\scriptstyle{u=(3,3,2,2)}$}
\put(256,-30){$\scriptstyle{v=0}$}

\put(300,-20){\line(1,-1){20}}
\put(340,-20){\line(-1,1){20}}
\qbezier[15](300,-20)(310,-10)(320,0)
\qbezier[15](320,-40)(330,-30)(340,-20)
\put(304,-9){$\scriptstyle{+}$}
\put(304,-34){$\scriptstyle{-}$}
\put(331,-34){$\scriptstyle{+}$}
\put(331,-9){$\scriptstyle{+}$}
\put(313,-22){$\scriptstyle{\sigma=24}$}
\put(345,-17){$\scriptstyle{u=(0,0,3,3)}$}
\put(356,-30){$\scriptstyle{v=1}$}

\put(0,-80){\line(1,1){20}}
\put(40,-80){\line(-1,-1){20}}
\qbezier[15](0,-80)(10,-90)(20,-100)
\qbezier[15](20,-60)(30,-70)(40,-80)
\put(4,-69){$\scriptstyle{+}$}
\put(4,-94){$\scriptstyle{+}$}
\put(31,-94){$\scriptstyle{-}$}
\put(31,-69){$\scriptstyle{-}$}
\put(13,-82){$\scriptstyle{\sigma=25}$}
\put(45,-77){$\scriptstyle{u=(1,2,3,0)}$}
\put(56,-90){$\scriptstyle{v=0}$}

\put(100,-80){\line(1,-1){20}}
\put(140,-80){\line(-1,1){20}}
\qbezier[15](100,-80)(110,-70)(120,-60)
\qbezier[15](120,-100)(130,-90)(140,-80)
\put(104,-69){$\scriptstyle{+}$}
\put(104,-94){$\scriptstyle{-}$}
\put(131,-94){$\scriptstyle{-}$}
\put(131,-69){$\scriptstyle{+}$}
\put(113,-82){$\scriptstyle{\sigma=26}$}
\put(145,-77){$\scriptstyle{u=(2,3,0,1)}$}
\put(156,-90){$\scriptstyle{v=1}$}

\put(200,-80){\line(1,1){20}}
\put(240,-80){\line(-1,-1){20}}
\qbezier[15](200,-80)(210,-90)(220,-100)
\qbezier[15](220,-60)(230,-70)(240,-80)
\put(204,-69){$\scriptstyle{-}$}
\put(204,-94){$\scriptstyle{-}$}
\put(231,-94){$\scriptstyle{+}$}
\put(231,-69){$\scriptstyle{+}$}
\put(213,-82){$\scriptstyle{\sigma=27}$}
\put(245,-77){$\scriptstyle{u=(3,0,1,2)}$}
\put(256,-90){$\scriptstyle{v=0}$}

\put(300,-80){\line(1,-1){20}}
\put(340,-80){\line(-1,1){20}}
\qbezier[15](300,-80)(310,-70)(320,-60)
\qbezier[15](320,-100)(330,-90)(340,-80)
\put(304,-69){$\scriptstyle{-}$}
\put(304,-94){$\scriptstyle{+}$}
\put(331,-94){$\scriptstyle{+}$}
\put(331,-69){$\scriptstyle{-}$}
\put(313,-82){$\scriptstyle{\sigma=28}$}
\put(345,-77){$\scriptstyle{u=(0,1,2,3)}$}
\put(356,-90){$\scriptstyle{v=1}$}

\put(0,-140){\line(1,-1){20}}
\put(40,-140){\line(-1,1){20}}
\qbezier[15](0,-140)(10,-130)(20,-120)
\qbezier[15](20,-160)(30,-150)(40,-140)
\put(4,-129){$\scriptstyle{+}$}
\put(4,-154){$\scriptstyle{+}$}
\put(31,-154){$\scriptstyle{-}$}
\put(31,-129){$\scriptstyle{-}$}
\put(13,-142){$\scriptstyle{\sigma=29}$}
\put(45,-137){$\scriptstyle{u=(3,2,1,0)}$}
\put(56,-150){$\scriptstyle{v=1}$}

\put(100,-140){\line(1,1){20}}
\put(140,-140){\line(-1,-1){20}}
\qbezier[15](100,-140)(110,-150)(120,-160)
\qbezier[15](120,-120)(130,-130)(140,-140)
\put(104,-129){$\scriptstyle{+}$}
\put(104,-154){$\scriptstyle{-}$}
\put(131,-154){$\scriptstyle{-}$}
\put(131,-129){$\scriptstyle{+}$}
\put(113,-142){$\scriptstyle{\sigma=30}$}
\put(145,-137){$\scriptstyle{u=(0,3,2,1)}$}
\put(156,-150){$\scriptstyle{v=0}$}

\put(200,-140){\line(1,-1){20}}
\put(240,-140){\line(-1,1){20}}
\qbezier[15](200,-140)(210,-130)(220,-120)
\qbezier[15](220,-160)(230,-150)(240,-140)
\put(204,-129){$\scriptstyle{-}$}
\put(204,-154){$\scriptstyle{-}$}
\put(231,-154){$\scriptstyle{+}$}
\put(231,-129){$\scriptstyle{+}$}
\put(213,-142){$\scriptstyle{\sigma=31}$}
\put(245,-137){$\scriptstyle{u=(1,0,3,2)}$}
\put(256,-150){$\scriptstyle{v=1}$}

\put(300,-140){\line(1,1){20}}
\put(340,-140){\line(-1,-1){20}}
\qbezier[15](300,-140)(310,-150)(320,-160)
\qbezier[15](320,-120)(330,-130)(340,-140)
\put(304,-129){$\scriptstyle{-}$}
\put(304,-154){$\scriptstyle{+}$}
\put(331,-154){$\scriptstyle{+}$}
\put(331,-129){$\scriptstyle{-}$}
\put(313,-142){$\scriptstyle{\sigma=32}$}
\put(345,-137){$\scriptstyle{u=(2,1,0,3)}$}
\put(356,-150){$\scriptstyle{v=0}$}

\end{picture} 
\end{center} 
\vskip 1.5 cm 
\caption{A possible representation of the vertex variables.
Each vertex is labelled by a variable $\sigma\in\{1,...,32\}$. 
Following \cite{Di2} a vector $u=(u_1,u_2,u_3,u_4)\in\{0,1,2,3\}^4$ and 
a scalar $v\in\{0,1\}$ are associated to any $\sigma\in\{1,\dots,32\}$
(see \cite{Di1} for more details).} 
\label{fig:quattro} 
\end{figure} 
 
\newpage 
\setlength{\unitlength}{2pt} 
\begin{figure} 
\begin{center} 
\begin{picture}(200,200)(0,-70) 
\thinlines 

\multiput(0,120)(0,40){3}{\line(1,0){80}} 
\multiput(0,120)(40,0){3}{\line(0,1){80}} 
\thicklines
\qbezier[15](0,140)(10,130)(20,120)
\qbezier[15](60,200)(70,190)(80,180)
\qbezier[45](20,200)(50,170)(80,140)
\qbezier[45](0,180)(30,150)(60,120)
\put(60,120){\line(1,1){20}}
\put(20,120){\line(1,1){60}}
\put(0,140){\line(1,1){60}}
\put(0,180){\line(1,1){20}}
\put(37,110){$(a)$}
\put(11,187){$-$}
\put(31,167){$+$}
\put(51,147){$-$}
\put(71,127){$+$}

\thinlines
\multiput(120,120)(0,40){3}{\line(1,0){80}} 
\multiput(120,120)(40,0){3}{\line(0,1){80}} 
\thicklines
\qbezier[15](120,140)(130,130)(140,120)
\qbezier[15](180,200)(190,190)(200,180)
\qbezier[45](140,200)(170,170)(200,140)
\qbezier[45](120,180)(150,150)(180,120)
\put(180,120){\line(1,1){20}}
\put(140,120){\line(1,1){60}}
\put(120,140){\line(1,1){60}}
\put(120,180){\line(1,1){20}}
\put(157,110){$(b)$}
\put(131,187){$-$}
\put(151,167){$-$}
\put(171,147){$-$}
\put(191,127){$-$}

\thinlines
\multiput(0,0)(0,40){3}{\line(1,0){80}} 
\multiput(0,0)(40,0){3}{\line(0,1){80}} 
\thicklines
\qbezier[15](20,0)(30,10)(40,20)
\qbezier[15](0,20)(10,30)(20,40)
\qbezier[15](0,60)(10,50)(20,40)
\qbezier[15](20,80)(30,70)(40,60)
\qbezier[15](40,20)(50,10)(60,0)
\qbezier[15](60,40)(70,30)(80,20)
\qbezier[15](60,40)(70,50)(80,60)
\qbezier[15](40,60)(50,70)(60,80)
\multiput(20,0)(20,20){4}{\line(-1,1){20}}
\multiput(0,60)(20,-20){4}{\line(1,1){20}}
\put(37,-10){$(c)$}
\put(7,72){$+$}
\put(31,47){$+$}
\put(71,7){$+$}
\put(5,7){$+$}
\put(71,71){$+$}

\thinlines
\multiput(120,0)(0,40){3}{\line(1,0){80}} 
\multiput(120,0)(40,0){3}{\line(0,1){80}} 
\thicklines
\qbezier[15](140,0)(150,10)(160,20)
\qbezier[15](120,20)(130,30)(140,40)
\qbezier[15](120,60)(130,50)(140,40)
\qbezier[15](140,80)(150,70)(160,60)
\qbezier[15](160,20)(170,10)(180,0)
\qbezier[15](180,40)(190,30)(200,20)
\qbezier[15](180,40)(190,50)(200,60)
\qbezier[15](160,60)(170,70)(180,80)
\multiput(140,0)(20,20){4}{\line(-1,1){20}}
\multiput(120,60)(20,-20){4}{\line(1,1){20}}
\put(157,-10){$(d)$}
\put(127,72){$+$}
\put(151,47){$-$}
\put(191,7){$+$}
\put(125,7){$+$}
\put(191,71){$+$}
\end{picture} 
\end{center} 
\vskip -0.5 cm 
\caption{The ground states depicted in figures ($a$), ($b$), ($c$) and
($d$) correspond, respectively, to the phases {\bf Fl}, 
{\bf SF}, {\bf F} and {\bf LF} in Fig. \ref{fig:cinque}.}   
\label{fig:tre} 
\end{figure} 

\newpage 
\begin{figure} 
\begin{picture}(200,200)(-80,-50)  
\multiput(0,0)(40,0){3}{B}
\multiput(0,40)(40,0){3}{B}
\multiput(0,80)(40,0){3}{B}
\multiput(20,0)(40,0){2}{A}
\multiput(20,40)(40,0){2}{A}
\multiput(20,80)(40,0){2}{A}
\multiput(0,20)(40,0){3}{D}
\multiput(0,60)(40,0){3}{D}
\multiput(20,20)(40,0){2}{C}
\multiput(20,60)(40,0){2}{C}
\thicklines
\put(21.5,25){\line(0,1){14}}
\put(21.5,45){\line(0,1){14}}
\put(41.5,25){\line(0,1){14}}
\put(41.5,45){\line(0,1){14}}
\put(24.5,42){\line(1,0){14}}
\put(44.5,42){\line(1,0){14}}
\put(24.5,62){\line(1,0){14}}
\put(44.5,62){\line(1,0){14}}
\thinlines 
\put(31,-10){\vector(0,1){40}}
\put(25,-15){$\rh{ABCD}$}
\put(90,32){\vector(-1,0){40}}
\put(95,32){$\rh{BADC}$}
\put(51,90){\vector(0,-1){40}}
\put(45,95){$\rh{DCBA}$}
\put(-10,52){\vector(1,0){40}}
\put(-25,52){$\rh{CDAB}$}
\end{picture} 
\vskip -2 cm  
\caption{Partition of the lattice used in the CVM calculation. 
The eight bonds and the four plaquettes are explicitely depicted.} 
\label{fig:reticolo} 
\end{figure}
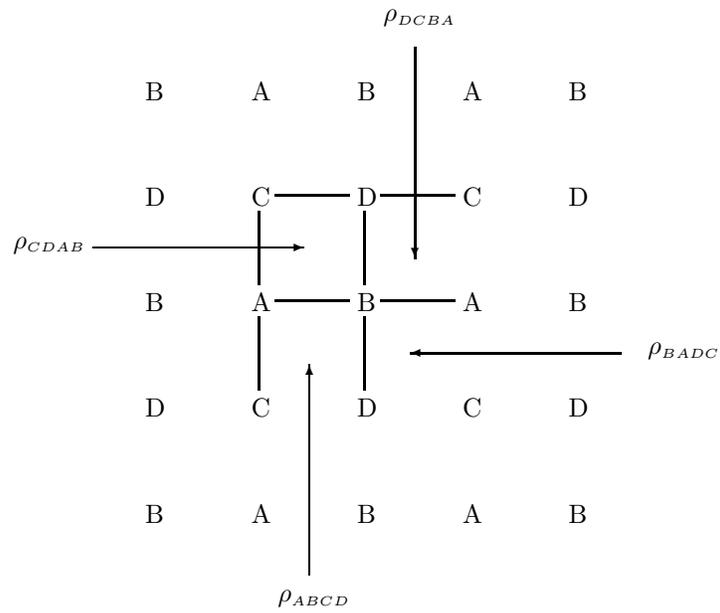 

\newpage

\end{document}